\documentclass[12pt,preprint]{aastex}
\begin{document}

\title{The Parsec-scale Accretion Disk in NGC\,3393}

\author{Paul T. Kondratko, Lincoln J. Greenhill, James M. Moran}

\affil{Harvard-Smithsonian Center for Astrophysics, 60 Garden St., Cambridge, MA
02138, USA}

\email{greenhill@cfa.harvard.edu}

\begin{abstract}
We present a Very Long Baseline Interferometry image of the water maser emission in the nuclear region of NGC\,3393. The maser emission has a linear distribution oriented at a position angle of $\sim -34\degr$, perpendicular to both the kpc-scale radio jet and the axis of the narrow line region. The position-velocity diagram displays a red-blue asymmetry about the systemic velocity and the estimated dynamical center, and is thus consistent with rotation. Assuming Keplerian rotation in an edge-on disk, we obtain an enclosed mass of $(3.1\pm 0.2) \times 10^7\,M_{\sun}$ within $0.36\pm 0.02$\,pc ($1.48\pm 0.06$\,mas), which corresponds to a mean mass density of $\sim10^{8.2}\,M_{\sun}$\,pc$^{-3}$. We also report the measurement with the Green Bank Telescope of a velocity drift, a manifestation of centripetal acceleration within the disk, of $5\pm 1$ km s $^{-1}$ yr$^{-1}$ in the $\sim3880$ km s$^{-1}$ maser feature, which is most likely located along the line of sight to the dynamical center of the system. From the acceleration of this feature, we estimate a disk radius of $0.17\pm 0.02$ pc, which is smaller than the inner disk radius ($0.36\pm 0.02$ pc) of emission that occurs along the midline (i.e., the line of nodes). The emission along the line of sight to the dynamical center evidently occurs much closer to the center than the emission from the disk midline, contrary to the situation in the archetypal maser systems NGC\,4258 and NGC1068.  The outer radius of the disk as traced by the masers along the midline is about 1.5 pc.
\end{abstract}


\section{Introduction}
\label{introduction}

NGC\,3393 is a face-on early spiral (Sa) galaxy classified optically as a Seyfert\,2 \citep{Veron2003}. It displays a prominent ring and interacts weakly with a companion $60$\,kpc away \citep{Schmitt2001}. There is compelling evidence that
the galaxy contains an active galactic nucleus (AGN). Its X-ray spectrum exhibits a $6.4$\,keV Fe K$\alpha$ line --- is an unambiguous indicator of nuclear activity \citep{Maiolino1998, Guainazzi2005, Levenson2006}.
\cite{Levenson2006} estimate the AGN's $2-10$\,keV intrinsic luminosity to be $10^{9.3}\,L_{\odot}$ based on modelling of the Fe line in the Chandra X-ray
Observatory data. A Compton-thick, cold reflection model applied to the BeppoSax
X-ray data yields an observed $2-10$\,keV flux of $3.9\times
10^{-13}$\,erg\,cm$^{-2}$\,s$^{-1}$ absorbed by a column density of
$>10^{25}$\,cm$^{-2}$ \citep{Maiolino1998}. If we assume, following
\cite{Comastri2004}, that the observed $2-10$\,keV flux is $1-10\%$ of the
unabsorbed $2-10$\,keV flux due to reflection and scattering, then the intrinsic
$2-10$\,keV luminosity is $10^{8.5-9.5}\,L_{\odot}$, which is consistent with the luminosity determined from the Fe K$\alpha$ line. Based on XMM and BeppoSax data, \cite{Guainazzi2005} report an absorbing column density of
$\left(4.4^{+2.5}_{-1.1}\right)\times 10^{24}$\,cm$^{-2}$, a spectral or photon of index $2.8^{+1.2}_{-0.7}$, and an observed flux of $\left(9^{+6}_{-4}\right)\times
10^{-14}$\,erg\,cm$^{-2}$\,s$^{-1}$, which corresponds to an intrinsic
$2-10$\,keV luminosity of $10^{10.2^{+2.0}_{-1.1}}\,L_{\odot}$ (as computed by the Portable Interactive Multi-Mission Simulator\footnote{http://heasarc.gsfc.nasa.gov/Tools/w3pimms.html}). In the
context of the AGN unified model, such a high column density is indicative of an
almost edge-on obscuring structure along the line of sight to the nucleus
\citep{Lawrence1982, Antonucci1993}. Furthermore, images of the inner-kpc region reveal outflows from the nucleus, consistent with the presence of an AGN. The Narrow Line Region (NLR) as traced by [O\,III] emission has an S-shaped morphology with an opening angle of $\sim90\degr$ and extends $\sim400$\,pc on either side of the nucleus along position angle (P.A.) of $\sim55^{\degr}$ \citep{Schmitt1996, Cooke2000}. Very Large Array (VLA) and Australia Telescope Compact Array (ATCA) observations reveal a double-sided jet with a total extent of $\sim700$\,pc along P.A.$\sim56\degr$ \citep{Schmitt2001b} or P.A.$\sim45^{\degr}$ \citep{Morganti1999}.
The existence of an AGN in the nucleus of NGC\,3393 is thus well established based on the available data.

NGC\,3393 does not show evidence of significant nuclear star formation. The IRAS
fluxes measured on an $\sim30$\,kpc scale \citep[assuming IRAS beam of
$\sim2'$;][]{Moshir1990} are consistent with a total infrared luminosity of
$10^{10}\,L_{\odot}$, a dust mass of $\sim5\times 10^{5}\,M_{\odot}$, and two
relatively cool dust components at $30$ and $130$\,K \citep{Diaz1988}. Under the
assumption that star formation alone is responsible for the infrared flux, the IRAS
luminosity yields a star formation rate of $\sim4$ $M_{\sun}$\,yr$^{-1}$ on kpc
scales \citep{Veilleux1994}, which is not consistent with a high level of
star-forming activity. Detailed stellar population synthesis work by
\cite{Fernandes2004} confirms that the blue optical spectrum of the central
$\sim200$\,pc is dominated by an old stellar population ($>2.5\times 10^9$\,yrs)
with a small contribution ($14\%$) to the featureless continuum from an AGN. Hence, starburst activity probably does not play a significant role in the nucleus of NGC\,3393.

The NGC\,3393 nucleus is also a source of water maser emission, which is
currently the only resolvable tracer of warm dense molecular gas in the inner
parsec of any AGN beyond $\sim1$\,Mpc. The maser spectrum of NGC\,3393 shows a
characteristic spectral signature of rotation in an edge-on disk: two complexes
($\sim70$\,mJy) symmetrically offset by $\sim600$\,km\,s$^{-1}$ from the
systemic velocity (henceforth, high-velocity emission) and a single spectral
complex ($\sim28$\,mJy) within $130$\,km\,s$^{-1}$ of the systemic velocity
\citep[henceforth, low-velocity emission;][]{Kondratko2006}.  Very Long Baseline Interferometry (VLBI) maps of seven water maser sources that show similar spectral signatures have been interpreted in a context of a model in which the maser emission traces a nearly edge-on disk of molecular material $0.1$ to $1$ pc from a supermassive black hole: NGC\,4258 \citep{Miyoshi1995, Herrnstein2005, Humphreys2008}, NGC\,1386 \citep{Braatz1997AAS}, NGC\,4945 \citep{Greenhill1997}, NGC\,1068 \citep{Greenhill1997c}, NGC\,3079 \citep{Trotter1998, Yamauchi2004, Kondratko2005}, IC\,2560 \citep{Ishihara2001}, Circinus \citep{Greenhill2003}. The maser spectrum of NGC\,3393 thus provides indirect evidence for an edge-on pc-scale molecular disk orbiting a supermassive black hole.

VLBI maps of nuclear water maser emission have been used to accurately estimate
black hole masses and pc-scale accretion disk geometries. In three water maser
systems mapped with the VLBI --- NGC\,4258 \citep{Miyoshi1995}, NGC\,1068 \citep{Greenhill1997c}, and the Circinus Galaxy \citep{Greenhill2003} --- position and line-of-sight velocity data provided evidence for differential rotation and enabled accurate estimation of black hole mass and pc-scale molecular disk structure. In another system, NGC\,3079, the rotation curve traced by the maser emission appears flat and was interpreted in the context of a pc-scale, thick, edge-on, self-gravitating, and possibly star forming molecular disk \citep{Kondratko2005}. In addition to mapping pc-scale molecular disk structure and accurately determining the mass of supermassive black holes, nuclear water maser emission has also been used as a distance indicator. Distance determination is possible for systems where a detailed knowledge of the structure of the disk from VLBI is combined with a measurement of either maser proper motions or drifts in line-of-sight velocity of spectral features (i.e., centripetal acceleration). The distance to NGC\,4258 obtained in this manner is the most accurate extragalactic distance thus far, is independent of standard candle calibrators such as Cepheids \citep{Herrnstein1999}, and has contributed to the analysis of the Cepheid period-luminosity relation \citep[Freedman et al. 2001;][]{Newman2001, Macri2007}. 

In this work, we present a VLBI map of the pc-scale accretion disk as traced by
water maser emission and estimate the mass of the black hole in NGC\,3393. Data
calibration and reduction techniques are discussed in Section \ref{observation}.
Spectral-line images of the inner-pc region are presented in Section \ref{results}. In Section \ref{discussion}, we interpret the observed kinematics of the maser distribution in terms of a nearly edge-on pc-scale molecular accretion disk that orbits a central mass. In this work, we adopt systemic velocity for NGC\,3393 based on $21$\,cm line measurements of $3750 \pm 5$\,km\,s$^{-1}$ \citep{Theureau1998}, which yields a distance of $50$\,Mpc for a Hubble constant of $75$\,km\,s$^{-1}$\,Mpc$^{-1}$. With this adopted distance, $1$\,mas corresponds to $0.24$\,pc.

\section{Observations and Calibration}
\label{observation} NGC\,3393 was observed in spectral-line mode with the Very
Long Baseline Array (VLBA) of the NRAO\footnote{The National Radio Astronomy
Observatory is operated by Associated Universities, Inc., under cooperative
agreement with the National Science Foundation}, augmented by the phased VLA and
the Green Bank Telescope (GBT), for approximately 5\,hours on 2004 April 26/27.
The source was observed with $8\times 16$\,MHz intermediate-frequency (IF) bands
that covered an uninterrupted velocity range of $3048$ to $4430$\,km\,s$^{-1}$.
Limits on the VLBA correlator data rate necessitated two processing passes through the correlator (with
$256$ channels per IF) and resulted in two separate data sets that do not overlap in velocity: one in left-circular polarization covering the velocity range from $3048$ to $3737$\,km\,s$^{-1}$, the other in right-circular polarization extending from $3737$ to $4430$\,km\,s$^{-1}$. The data were reduced using the Astronomical Image Processing System. The amplitude calibration included corrections for atmospheric opacity. Antenna gain curves and measurements of system temperature were used to calibrate the amplitude data for each of the VLBA stations. Amplitude calibration for the VLA was based on the measurements with the VLA of flux densities for VLBI calibrators with respect to 3C\,286, for which we adopted a $22$-GHz flux density of $2.56$\,Jy. To minimize systematic errors due to atmospheric opacity effects, we used in this calibration only VLA scans of VLBI calibrators with elevations close to the elevation of 3C\,286 ($elev=28\degr$, $29\degr$, $54\degr$, $43\degr$, $64\degr$ for 3C\,286, 1055-248, 4C\,39.25, 0727-115, 0748+126, respectively).

The data in each polarization were corrected for the effect of the parallactic angle of the source on fringe phase. The zenith path delay over each antenna was estimated via observations of $10$ strong ($>0.5$\,Jy) calibrators with sub-mas positions for $\sim1$\,hour before and $\sim1$\,hour after the main imaging experiment. The residual multiband delays after applying this calibration were $<0.2$\,ns, which corresponds to a phase error of less than $1\degr$ for a $16$\,MHz IF band (see Table \ref{error}). Time variation in delays and fringe rates due to the troposphere and clock uncertainties as well as electronic phase difference among bands were removed via $\sim4$\,min observations of strong ($3-7$\,Jy) calibrators (0727-115, 0748+126, 4C\,39.25) made approximately every hour. We estimate that this calibration is accurate to within $0.2$\,ns and $1$\,mHz for delays and rates, respectively; we note that a residual rate of $1$\,mHz results in a phase error of $0.5\degr$ (if not corrected) over a $1$\,hour time span, the approximate time separation between the calibrator scans (Table \ref{error}). The complex bandpass shapes of each antenna were corrected using the data on 0727-115, 0748+126, and 4C\,39.25. Phase and amplitude fluctuations due to the troposphere and clock uncertainties were removed via observations every $\sim100$s of 1055-248, located $2.4\degr$ away from NGC\,3393, self-calibrating the data on 1055-248, and applying the resulting phase and amplitude solutions to the target source. As a result of this process, the position of the maser on the sky was determined relative to 1055-248. Based on VLA data obtained at two epochs in B and BnA configuration, we estimate the $22$\,GHz flux of 1055-248 to be $0.38 \pm 0.03$\,Jy. This measurement as well as the photometric data from the NASA extragalactic database (NED) and from NRAO VLA Sky Survey \citep[$0.33$\,Jy at $1.4$\,GHz;][]{Condon1998} suggests that 1055-248 is a flat spectrum source and is therefore expected to be unresolved on mas scales. In fact, using the source model obtained from self-calibration, we estimate that $90\%$ of 1055-248 flux originates from a point source. The calibrator appears nearly point-like also in a narrow bandwidth (250\,kHz $\sim3.4$\,km\,s$^{-1}$), which confirms the quality of bandpass calibration. Based on the consideration of the tropospheric structure function as predicted by the Kolmogorov turbulence theory \citep{Carilli1999}, we estimate that the interferometer phase towards 1055-248 reproduces that towards NGC\,3393 to within $\phi_{\rm rms}<40\degr$, (assuming a water vapor scale height of $\sim2$\,km, a representative antenna elevation angle of $\sim25^{\degr}$, and a tropospheric pattern speed of $<15$\,m\,s$^{-1}$; Table \ref{error}) which corresponds to a loss of signal on the target source of at most $1-\exp[-\phi_{\rm rms}^2/2]=0.22$. The uncertainty in phase due to calibrator-target angular separation is the most significant source of error in this experiment and depends sensitively on the assumed pattern speed, which is not well constrained for this experiment (Table \ref{error}).

After calibrating and imaging the spectral-line data set, the positions and the
integrated fluxes of the maser features were obtained by fitting two-dimensional
elliptical Gaussians to the distribution of the maser emission on the sky in each spectral channel. The resulting uncertainties in maser spot positions, based on formal error analysis, are consistent with the theoretical expectations based on signal-to-noise ratio (SNR) and beam size. The synthesized beam dimensions were $3.5 \times 1$\,mas at P.A.$\sim -1\degr$ and the resulting noise level was $\sim5$ mJy in the $\sim3.4$\,km\,s$^{-1}$ spectral channels, a spectral breadth comparable to the linewidths of the maser features. To estimate the fraction of the power imaged by the interferometer and to monitor the maser with the intent of measuring velocity drifts, we acquired single-dish spectra of NGC\,3393 with the GBT using the configuration and calibration described in \cite{Kondratko2008}. All velocities have been computed in accordance with the optical definition of Doppler shift and are in the heliocentric reference frame.

\section{Results}
\label{results}

\begin{figure*}[!h]
\plotone{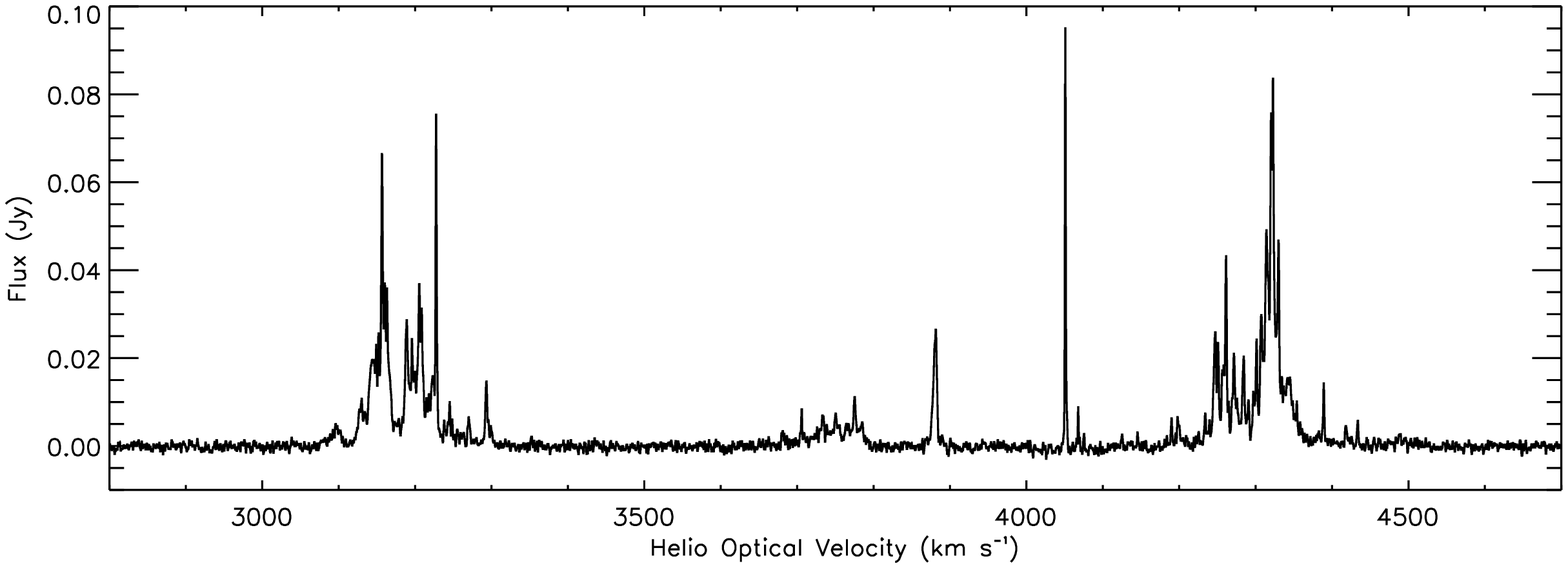} \plotone{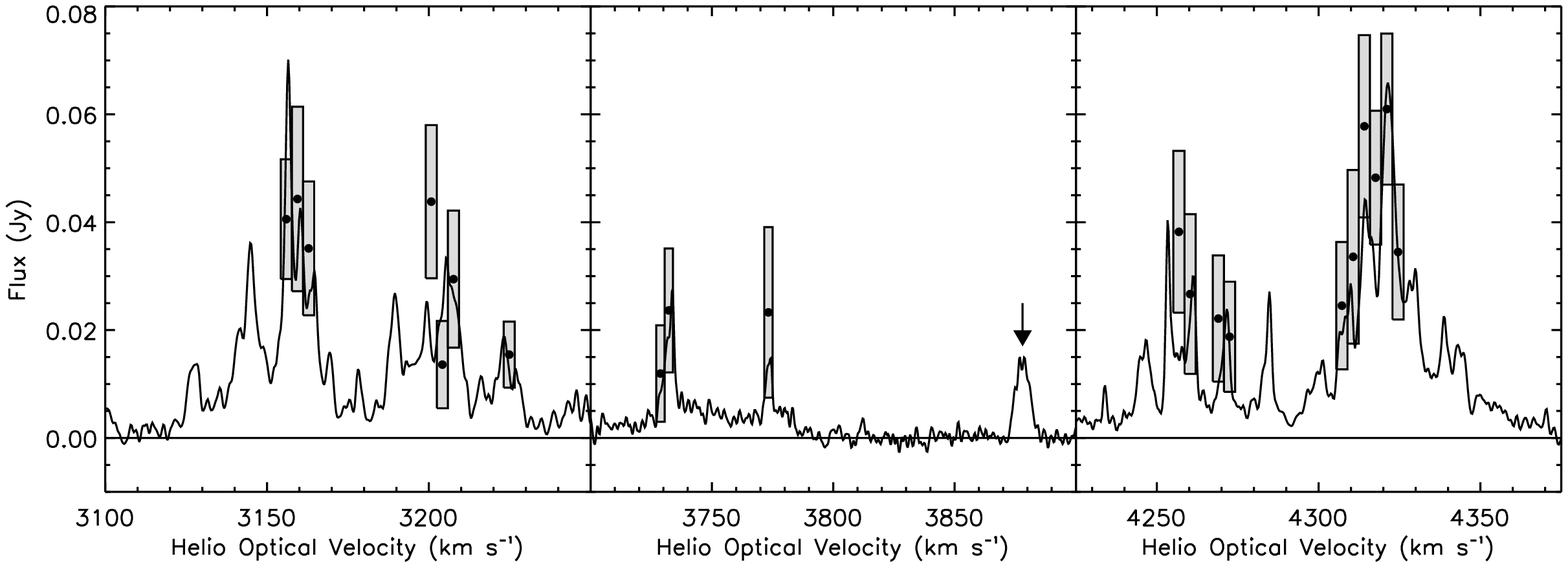}\caption{{\it Top:} A weighted average of five total-power spectra of the maser in NGC\,3393 obtained with the GBT on 2005 October 17, 2006 January 22, 2006 March 23, 2006 April 28, and 2006 May 23 with an effective resolution of $108\,\mbox{kHz}$ ($\sim1.5$\,km\,s$^{-1}$) and $1\sigma=0.6$\,mJy.
        {\it Bottom:} Comparison of VLBI imaged power and total power spectrum.  The total-power spectrum was obtained with the GBT on 2005 January 15, with an effective resolution of 108 kHz ($\sim 1.5$\,km\,s$^{-1}$).  Angle-integrated flux density (filled circles) and $1\sigma$ measurement uncertainty (shaded boxes) are shown for each $3.4$\,km\,s$^{-1}$ spectral channel with detectable emission.  Though separated by $\sim9$\,months, the single-dish and VLBI flux measurements agree in general to $\sim1\sigma$, the exception being the apparent absence of $3880$\,km\,s$^{-1}$ emission (marked by the arrow and for which we report a velocity drift of $5\pm1$\,km\,s$^{-1}$\,yr$^{-1}$) at the time of the VLBI observations.
        \label{spectrum}}
        \hrulefill\
\end{figure*}

\begin{figure*}[!h]
\plotone{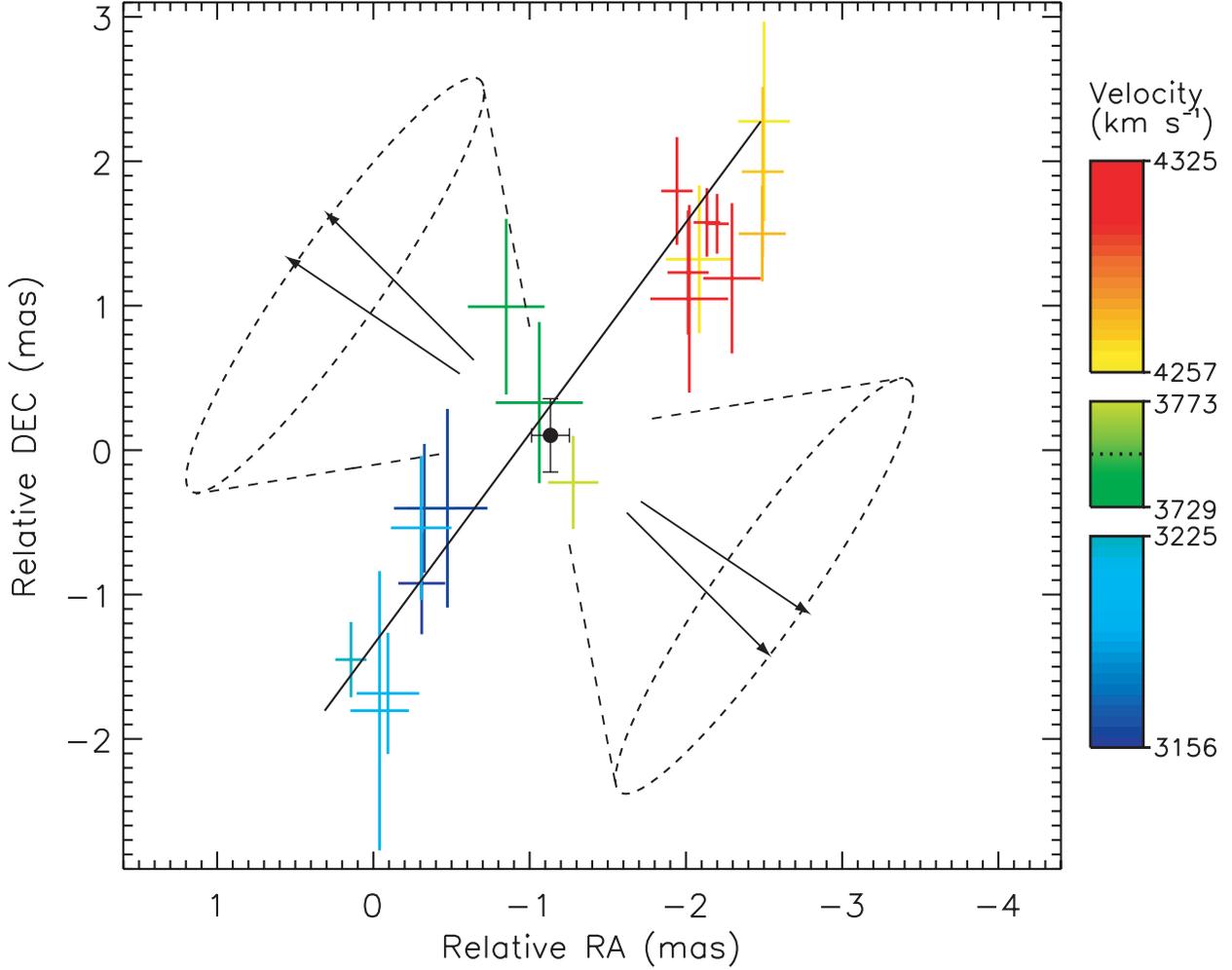} \caption{Distribution of maser emission in the nuclear region of NGC\,3393. Position uncertainties are $1\sigma$, and the colors of the maser spots indicate heliocentric optical line-of-sight velocity in accordance with the bar on the right. The dotted line in the color bar shows the adopted systemic velocity of $3750$\,km\,s$^{-1}$. The adopted location for the dynamical center (black circle) is the weighted mean for the low-velocity maser features. A line fitted to the distribution of maser emission on the sky (P.A.$\sim -34\degr$) is close to orthogonal to the kpc-scale radio jet (black arrows: P.A.$\sim 45\degr$, Morganti et al. 1999; $\sim 56\degr$, Schmitt et al. 2001b) and to the axis of the NLR \cite[dashed cone: P.A.$\sim55\degr$ with an opening angle of $\sim90\degr$][]{Schmitt1996, Cooke2000}. The coordinates are relative to $\alpha_{2000}=10^h48^m23\fs4660$ and $\delta_{2000}= -25\arcdeg09\arcmin43\farcs478$. At a distance of $50$\,Mpc, $0.24$ pc subtends $1$\,mas.
        \label{map}}
        \hrulefill\
\end{figure*}

The spectrum of imaged power agrees to within $2\sigma$ with the total power
spectrum obtained with the GBT about $9$ months after the VLBI observation (Fig.
\ref{spectrum} and Table \ref{maser_spots}). The difference between the two spectra may be due to source variability, since single-dish monitoring of water maser sources has revealed substantial flux variability on timescales of months to years (this work for NGC\,3393 and \cite{Baan1996} for other
sources). The maser emission is distributed on the sky in a
linear arrangement at P.A.$\sim -34\degr$ (Fig. \ref{map} and Table
\ref{maser_spots}), perpendicular to both the kpc-scale radio jet (P.A.$\sim
45\degr$, Morganti et al. 1999; $\sim56\degr$ Schmitt et al. 2001b)\nocite{Morganti1999, Schmitt2001b} and the axis of the NLR
\citep[P.A.$\sim55\degr$;][]{Schmitt1996, Cooke2000}. The maser
emission on the sky is clearly systematically distributed according to  velocity: emission red- and blue-shifted with respect to the systemic velocity are located in the
north-western and south-eastern parts of the image, respectively. Although the
two VLBA correlator passes did not overlap in velocity, we can nevertheless place a limit on their registration by aligning the common low velocity maser spots in the two data sets. If we assume that the mapped low-velocity features arise at the same location on the sky, then the two correlator passes are registered to within $0.3 \pm 0.1$ mas and $0.9 \pm 0.3$ mas in RA and DEC, respectively. The
systematic offset between the results of the two correlator passes was also estimated by
applying the calibration of one data set to the other. The transfer of phase and
amplitude solutions from self-calibration on 1055-248 from one correlator pass to the other resulted in an offset of 1055-248 from map center of $0.30 \pm 0.02$ mas and $0.52 \pm 0.06$ mas in RA and DEC, respectively, which is consistent with the offsets based on the location of the low-velocity maser features. These offsets provide an estimate of the systematic uncertainty in the registration of the two velocity sections of the maps.  Note that if we were to apply these offsets, the high velocity masers would better fit a straight line distribution, and the low velocity features would be more tightly clustered (see Figure \ref{map}).  However, this shift has an insignificant effect on the disk parameters derived in the next section.  Radio maps of the inner-kpc are suggestive of a jet perpendicular to the distribution of maser emission on the sky. However, we find no evidence in our data at a $4\sigma=2.4$\,mJy level for the $22$ GHz continuum on $\sim50$ pc scale.

\begin{figure*}[!h]
\plottwo{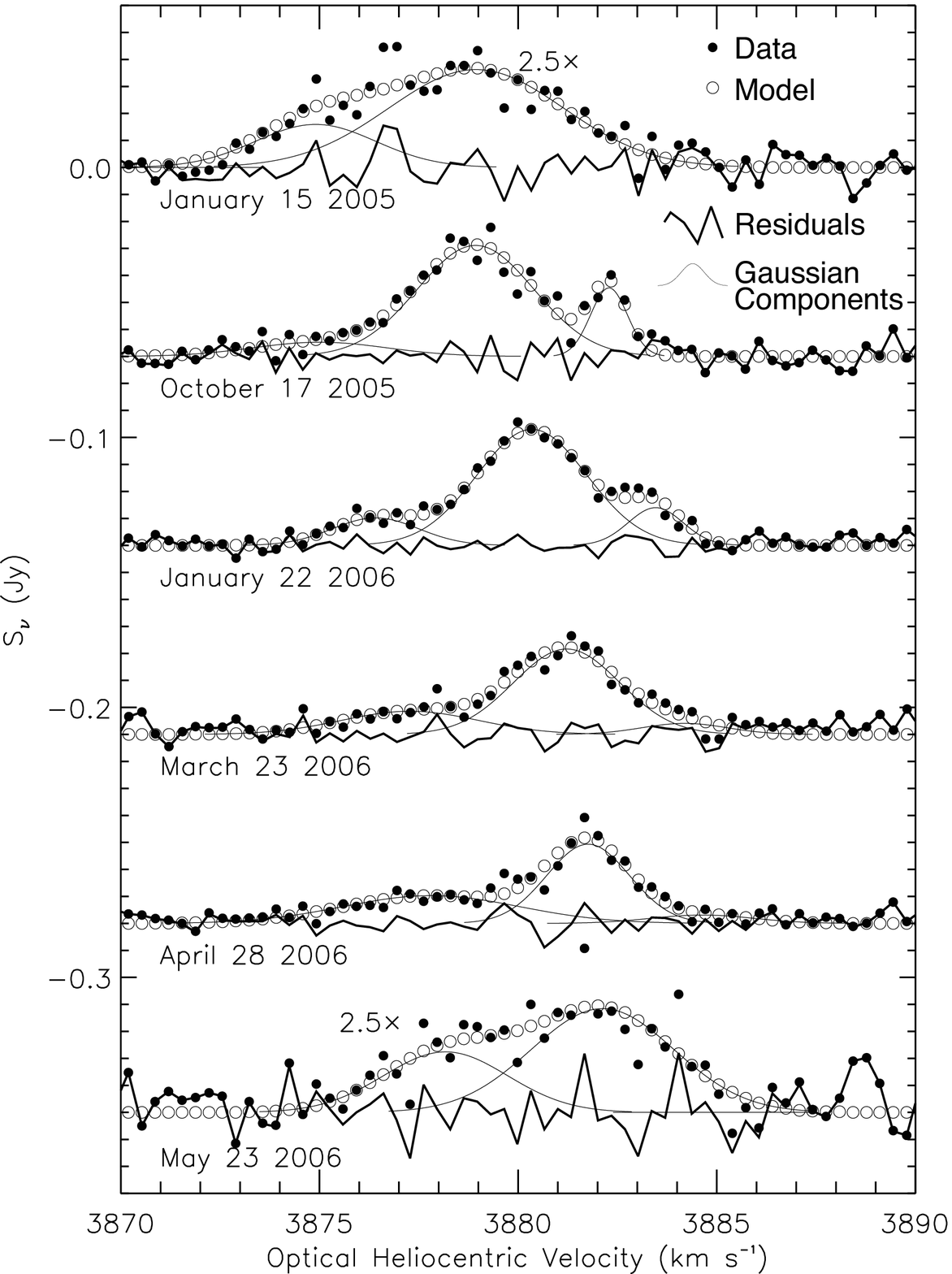}{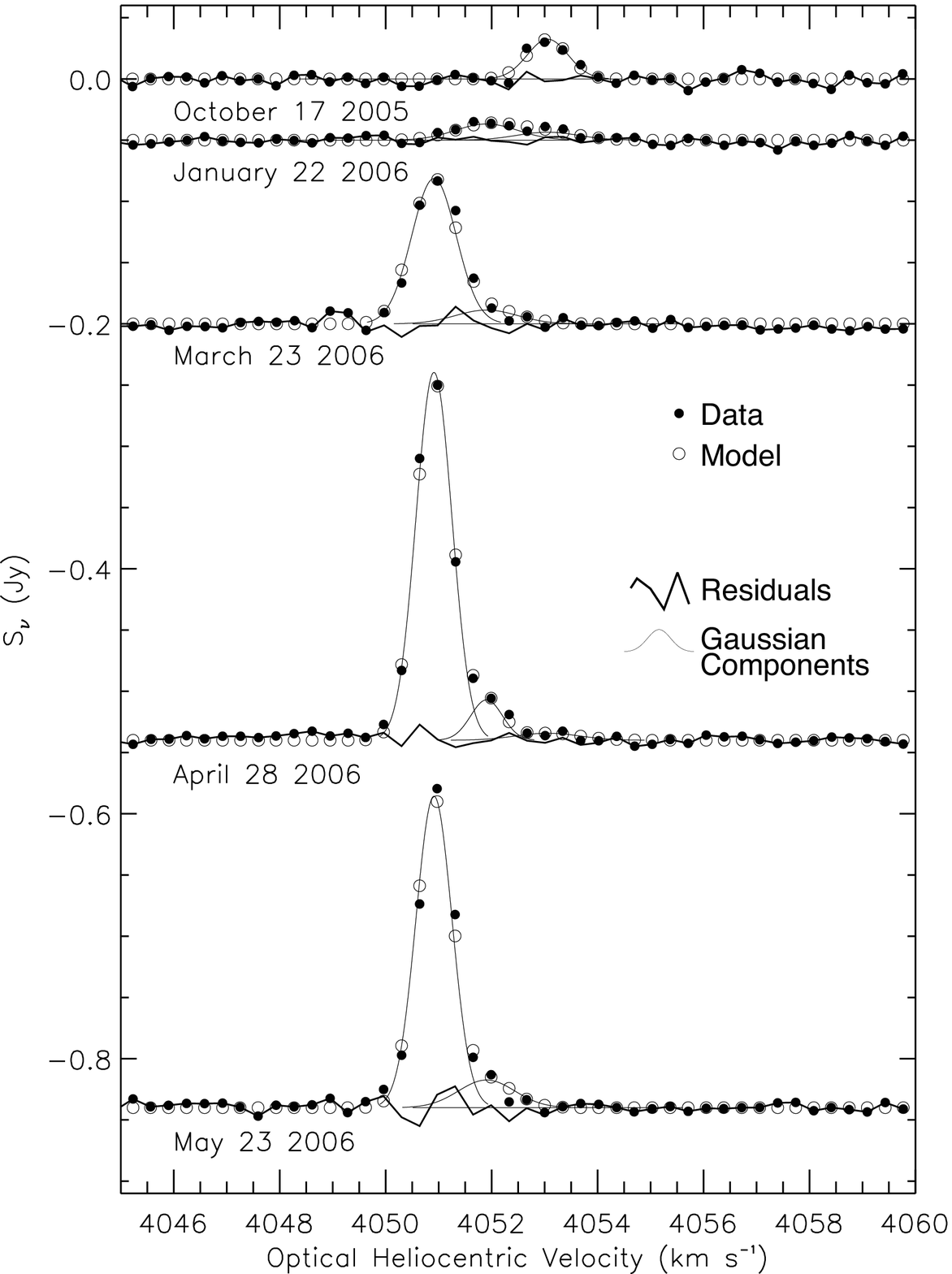} \caption{Maser spectra of features at
$\sim3880$ km s$^{-1}$ and $\sim4051$ km s$^{-1}$ obtained with the Green
Bank Telescope and the results of Gaussian component decomposition. The feature near $4051$ km s$^{-1}$ was not detected on 15 January 2005, emission in
the immediate vicinity of the systemic velocity, $\sim3750$ km s$^{-1}$, was
too weak to allow Gaussian component decomposition and is not shown.
\label{gauss_fit}}
        \hrulefill\
\end{figure*}

The low-velocity spectral features are expected to drift in velocity since they
are presumably moving across the line of sight where the centripetal acceleration vector is along the line of sight, and therefore at a maximum. To determine these velocity drifts, we monitored
NGC\,3393 with the GBT and report the measurement of the centripetal acceleration in the low-velocity complex at $\sim3880$\,km\,s$^{-1}$ (Fig. \ref{gauss_fit} and Table \ref{gauss_fit_table}). Using the code described in \cite{Humphreys2008}, we decomposed the spectra for the six epochs into three Gaussian components and used an iterative least squares technique to solve simultaneously for velocity drifts, amplitudes, peak centroids, and widths of all Gaussian components at all epochs. We obtain a reduced $\chi^2$ of $1.2$ and velocity drifts of $5.3 \pm 0.7$, $5.3 \pm 0.2$, and $4.4\pm 0.2$\,km\,s$^{-1}$\,yr$^{-1}$ for components at $3871.0 \pm 0.9$, $3874.9 \pm 0.2$, and $3879 \pm 0.2$ km s $^{-1}$, respectively, where the reference epoch for component velocities is 2005 January 15. In addition, spectra obtained since 2004 October 17 reveal a strong ($<0.3$\,Jy) feature at $\sim4051$\,km\,s$^{-1}$ that was present neither in earlier spectra of the source nor in the VLBI map (Fig. \ref{gauss_fit} and Table \ref{gauss_fit_table}). The first three epochs on this feature showed a negative centripetal acceleration of about $-4$\,km\,s$^{-1}$\,yr$^{-1}$, as would be expected for low-velocity emission that arises from behind the dynamical center; however, the line stopped drifting in the more recent spectra. We note that such behavior can be reproduced by a variation in strength of multiple components that do not drift in velocity (i.e., stationary components). In fact, from a Gaussian decomposition of the five available epochs, we infer that the data on this complex are consistent (reduced $\chi^2$ of $2.0$) with three time variable but stationary ($<0.7$\,km\,s$^{-1}$\,yr$^{-1}$) Gaussian components at $4051$, $4052$, and $4053$\,km\,s$^{-1}$. High-velocity emission is not expected to drift in velocity since it is located along the disk midline where the centripetal acceleration vector is perpendicular to the line of sight. We thus suggest that the newly detected complex at $\sim4051$\,km\,s$^{-1}$ may be a high-velocity emission component. We note that low-velocity emission that arises from behind the dynamical center has not been detected to date from any known nuclear water maser sources. This non-detection can be explained if free-free absorption by an
intervening ionized gas is considerable \citep[e.g.,][]{Herrnstein1996} or the
presence of a background nuclear continuum source is necessary to generate, via
amplification, low-velocity emission luminous enough for us to detect
\citep[e.g.,][]{Herrnstein1997}.

\section{Discussion}
\label{discussion} We interpret the linear distribution of the maser emission,
perpendicular to the radio jet and to the axis of the NLR, as well as the
segregation of the blue- and the red-shifted emission on the sky in the context
of a pc-scale molecular disk. We assume that the disk is nearly edge-on and,
based on the measured positive centripetal acceleration, we infer that the
low-velocity emission lies in front of and along the line of sight to the
dynamical center. The distribution of maser emission on the sky is consistent
with a relatively straight (i.e., nonwarped) disk. We note that a 4th degree polynomial fit to the maser distribution yields a marginal (i.e., $\sim10$\%) improvement in $\chi^2$ over a straight line fit. Thus, the evidence for a warp is at most tentative with our SNR.

We use the mean position of the low-velocity maser features to estimate the
location of the dynamical center (Fig. \ref{map}), RA$_{BH}$ and DEC$_{BH}$.  The absolute position of the dynamical center is 

$$\alpha_{2000} = 10^h48^m23\fs4659\pm0\fs0001$$
$$\delta_{2000} = -25\arcdeg09\arcmin43\farcs477\pm0\farcs001$$  

\noindent The errors are dominated by the uncertainty in the position of 1055-248 of about 1 mas \citep{Beasley2002}.  

The impact parameter of each maser feature was calculated as\\$\sqrt{(RA_i-RA_{BH})^2+(DEC_i-DEC_{BH})^2}$. The resulting position-velocity diagram displays a red-blue antisymmetry about the adopted $v_{sys}$ and estimated dynamical center and is thus consistent with
rotation (Fig. \ref{velocity}). From a fit of the Keplerian rotation to the
high-velocity features ($\chi^2_R=0.6$), we obtain a mass of $(3.1\pm 0.2)\times
10^7\,M_{\sun}$ enclosed within $0.36\pm0.01$\,pc ($1.48\pm0.06$\,mas; to
estimate the minimum impact parameter from high-velocity data, we used the
probability distribution function for a minimum of random variables following the Rice distribution). The fit to the blue- or the red-shifted emission alone yields $(3.0\pm 0.3)\times10^7\,M_{\sun}$ or $(3.2\pm 0.2)\times10^7\,M_{\sun}$,
respectively. If one correlator pass is shifted $0.9$\,mas so that its single
low-velocity maser feature overlaps the two low-velocity features in the other
data set, then the mass becomes $(3.5\pm 0.2)\times 10^7\,M_{\sun}$ enclosed
within $0.41\pm0.02$\,pc ($1.71\pm0.07$\,mas), which reflects the sensitivity of
our results to systematic errors. We note that, in addition to the Keplerian
rotation, the data are also consistent (in the sense that $\chi^2_R\lesssim 1$)
with $v\propto r^{\alpha}$ for $-1\le\alpha\le -0.1$. In particular, we obtain a
minimum of $\chi^2_R=0.4$ at $\alpha=-0.2$, which suggests that the disk might have significant mass with respect to that of the black hole.  The central mass would be $2.6\times 10^7\,M_{\odot}$ enclosed within $0.36$ pc (if we assume spherical symmetry). The mass of the disk traced by the high-velocity maser emission ($0.36-1$\,pc) is $1.9\times 10^7$\,$M_{\odot}$, which can be compared to disk masses (computed on similar scales) of $\sim7\times 10^{6}$ $M_{\odot}$ and $\sim9\times 10^{6}$ $M_{\odot}$ for NGC\,3079 \citep{Kondratko2005} and NGC\,1068 \citep{Lodato2003}, respectively. A
flat rotation curve model, which gives $1.1\times 10^7\,M_{\odot}$ enclosed
within $0.16$\,pc, is excluded by the data ($\chi^2_R\approx 20$), unless
velocity dispersion on the order of $30$\,km\,s$^{-1}$ is included. Such a large
velocity dispersion would most likely be indicative of macroscopic random motions among the molecular clumps responsible for the maser emission rather than turbulence within the clumps.

The mean mass density corresponding to $3.1\times 10^7$ $M_{\sun}$ enclosed
within $0.36$ pc is $\sim10^{8.2}\,M_{\sun}$ pc$^{-3}$. The relatively high
mean mass density for NGC\,3393 is suggestive of a massive central black hole, which is consistent with the X-ray observations of the nucleus. The estimated enclosed mass of $(3.1\pm 0.2)\times10^7$\,$M_{\sun}$ is in agreement with the empirical relation between bulge velocity dispersion and black hole mass
\citep[Gebhardt et al. 2000a, 2000b;][]{Ferrarese2000, Ferrarese2001}. If we
adopt $M_{BH}=1.2\times10^8\,M_{\odot}\,[\sigma/(200\mbox{ km s}^{-1})]^{3.75}$
(Gebhardt et al. 2000a, 2000b)\nocite{Gebhardt2000a}\nocite{Gebhardt2000b}, then
the velocity dispersion of the bulge, $184\pm 18$\,km\,s$^{-1}$ from the central
$\sim35$\,kpc \citep{Terlevich1990} or $157\pm 20$\,km\,s$^{-1}$ from the central $\sim200$\,pc \citep{Fernandes2004}, predicts black hole masses of $(9\pm 4)\times 10^7$\,$M_{\odot}$ or $(5\pm 2)\times 10^7$\,$M_{\odot}$, respectively, roughly a factor of two higher than our measurement. 

The most reliable estimate of the AGN's $2-10$\,keV intrinsic luminosity is
$10^{9.3}\,L_{\odot}$ determined from Fe K$\alpha$ line luminosity
\citep{Levenson2006}. Based on an average quasar spectral energy distribution
\citep{Fabian1999, Elvis2002}, the $2-10$\,keV luminosity is $1-3\%$ of the AGN bolometric luminosity and we obtain a bolometric luminosity for NGC\,3393 of
$10^{10.8-11.3}$\,$L_{\odot}$. We note that this estimate for the AGN bolometric luminosity is consistent with the total IRAS luminosity of the source of
$\sim10^{10}\,L_{\odot}$ \citep[measured on $\sim30$\,kpc scale;][]{Moshir1990}. The Eddington luminosity of a $3\times10^7$\,$M_{\odot}$ object is $10^{12}$\,$L_{\odot}$. Assuming that all of the enclosed mass is concentrated in a supermassive black hole, the bolometric luminosity of the central engine yields an Eddington ratio of $0.06-0.2$, which is consistent with the $0.01-1$ range obtained for Seyfert 1 galaxies, representative supermassive black hole systems \citep[e.g.,][]{Padovani1989, Wandel1999}, but larger than for advection dominated accretion flow systems \citep[e.g., $10^{-3.6\pm 1}$;][]{Herrnstein1998, Yuan2002}.  Assuming a standard
accretion efficiency of $\sim0.1$ (Frank, King, \& Raine 2002; see also Marconi
et al. 2004)\nocite{Frank2002}\nocite{Marconi2004}, we estimate a mass accretion
rate of $\dot{M}=0.04-0.1$\,$M_{\sun}$\,year$^{-1}$.

For a central mass of $(3.1\pm 0.2)\times 10^7\,M_{\sun}$ and the measured
centripetal acceleration of $a=5\pm 1$\,km\,s$^{-1}$\,yr$^{-1}$, we estimate disk
radius of the systemic maser feature at $\sim3880$\,km\,s$^{-1}$ of
$r_{sys}=\sqrt{GM_{BH}/a}=0.17\pm 0.02$\,pc, which is
significantly smaller than the inner disk radius of the high-velocity emission
($0.36\pm 0.02$\,pc). Evidently, the systemic emission in NGC\,3393 arises much
closer to the dynamical center than the high-velocity emission, which is in
contrast to the situation in NGC\,4258 \citep{Herrnstein2005} and NGC\,1068
\citep{Greenhill1997}, where disk radii of low-velocity features is about equal to the inner radius of the high velocity masers. It has been suggested that the
systemic emission in NGC\,4258 resides in a bowl that is a consequence of an
inclination-warped disk \citep{Herrnstein2005}. Such a warp in the accretion disk
structure might also determine the preferred radial location of the low-velocity
features in NGC\,3393. The resulting orbital velocity of the
$\sim3880$\,km\,s$^{-1}$ maser feature is $890\pm 60$\,km\,s$^{-1}$ (which might
be as high as $920\pm 60$\,km\,s$^{-1}$ due to systematic errors). We note that
the $130$\,km\,s$^{-1}$ offset of this feature from the adopted systemic velocity
might be due its location within the disk at a non-zero azimuthal angle $\phi$
from the line of sight to the central engine. Using this velocity offset, we
estimate $\phi=8\degr$ and the resulting corrections to the derived radius and
orbital velocity are much smaller than the corresponding uncertainties. The newly detected feature at $\sim4051$\,km\,s$^{-1}$ that we postulate to be
high-velocity emission was not detected in the VLBI experiment but would appear
at a large disk radius which from the computed enclosed mass and Keplerian formula, $r=GM_{BH}/v^2$, is $\sim6$\,mas or $1.5$\,pc (see Figure \ref{velocity}).  Hence, we estimate that the accretion
disk extends from $0.17$\,pc to $1.5$\,pc.

\section{Conclusion}

\begin{figure*}[!h]
\plotone{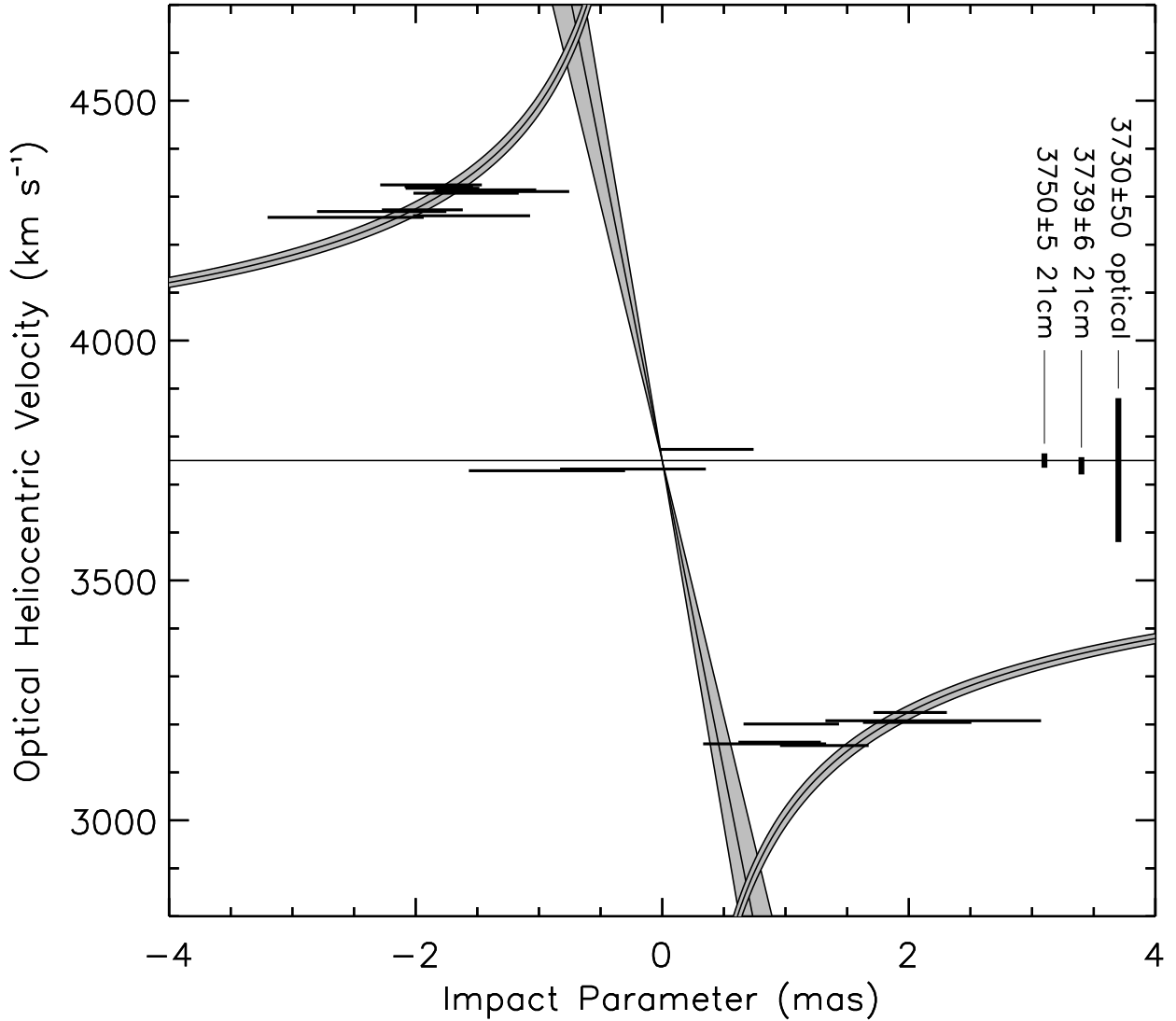} \caption{Position-velocity diagram for the detected maser 
emission in NGC\,3393. A fit of the Keplerian rotation law to the high-velocity
features (filled grey curves) yields a mass of $(3.1\pm 0.2)
\times10^7\,M_{\sun}$ enclosed within $0.36\pm 0.02$\,pc ($1.48\pm 0.06$\,mas).
From the measured velocity drift ($a=5\pm 1$\,km\,s$^{-1}$\,yr$^{-1}$), we infer
the radius of the systemic feature at $\sim3880$\,km\,s$^{-1}$ (not detected at the VLBI epoch itself)of
$r_{sys}=\sqrt{GM_{BH}/a}=0.17\pm0.02$\,pc, which corresponds to
$\Omega_{sys}=\sqrt{GM_{BH}/r_{sys}^3}= a^{3/4}/(GM_{BH})^{1/4}=0.005\pm
0.001$\,rad\,yr$^{-1}$ (nearly vertical, filled grey cone). The systemic velocity estimates are cited by the NED ($3\sigma$ uncertainties are plotted). In our calculations, we adopted the systemic velocity
from $21$\,cm line measurements of $3750\pm 5$\,km\,s$^{-1}$
\citep{Theureau1998}.\label{velocity}}
        \hrulefill\
\end{figure*}

We have mapped for the first time the maser emission in the nuclear region of
NGC\,3393. We interpret the linear distribution of the maser emission and the
segregation of the blue- and the red-shifted emission on the sky in the context
of a pc-scale nearly edge-on molecular disk that orbits a central mass of
$(3.1\pm 0.2)\times 10^7\,M_{\sun}$ enclosed within $0.36\pm 0.02$\,pc
($1.48\pm0.06 $\,mas). We also report the measurement of centripetal
acceleration, $a=5\pm 1$\,km\,s$^{-1}$\,yr$^{-1}$, in the low-velocity maser
feature at $\sim3880$\,km\,s$^{-1}$, which yields disk radius of
$0.17\pm0.02$\,pc for the derived central mass. The low-velocity emission in
NGC\,3393 occurs much closer to the dynamical center than the high-velocity
emission, in contrast to the situation in NGC\,4258 and NGC\,1068, two
archetypal maser systems. An independent estimate for the disk radius of the
low-velocity features would be provided by the measurement of their proper
motions. For a distance $D$, a central mass
$3.1\,(D/50\,\mbox{Mpc})\times10^7\,M_{\sun}$, and a radius
$0.17\,(D/50\,\mbox{Mpc})^{1/2}$\,pc, we expect motions of
$\sim4\,(D/50\,\mbox{Mpc})^{-3/4}\,\mu$as\,yr$^{-1}$, which would be challenging
to measure because of the typical lifetimes of the maser's features and their weakness. Alternatively, a measurement of the position-velocity gradient in the low-velocity maser features would provide an independent estimate for their radial location within the disk
($r_{sys}=0.17\,[D/50\,\mbox{Mpc}]\,[\Delta/0.27\,\mbox{Mpc}\,\mbox{yr}^{-1}\,\mbox{rad}^{-1}]^{-2/3}$\,pc,
where $D$ is the distance and $\Delta=v/\theta$ is the velocity
gradient).  The limited SNR in our VLBI data precluded measurement of the gradient (as evident from Figure \ref{velocity}). It is unclear what improvement in SNR would be necessary to yield a useful measurement, as there is a dearth of low-velocity features even in the sensitive single-dish spectra of the source.  Nonetheless, the maser is time variable, and new spectral features may emerge with time. An independent estimate for the disk radius of the low-velocity features either from proper motions or position-velocity gradient could be used to determine a distance to NGC\,3393, a result of considerable value since the galaxy is within the Hubble flow ($v_{\rm sys}=3750$ km\,s$^{-1}$) and thus might be used to establish a Hubble relation independent of standard candle calibrators such as Cepheids \cite[e.g.,][]{greenhill2004}.  Over and above eventual modeling errors for VLBI data, the peculiar motion of NGC3393 (or the barycenter of the parent Hydra cluster) would probably limit the accuracy of inference for the Hubble constant from NGC3393 alone to $\sim10\%$.  Within this, uncertainty over the flow field in the vicinity of the Great Attractor would probably dominate \citep{Masters2006}.

\acknowledgements
We thank M.\,Reid for suggestions related to VLBI scheduling and software that aided atmospheric delay calibration and component analysis of maser spectra. We also thank M.\,Elvis and R.\,Narayan for helpful discussions, C.\,Bignell for
flexibility in GBT scheduling, and J.\,Braatz for help in GBT set-up and
observing. This research has made extensive use of the NASA/IPAC Extragalactic
Database (NED) which is operated by the Jet Propulsion Laboratory (JPL),
California Institute of Technology, under contract with NASA. This work was
supported by GBT student support program, grants GSSP004-0005 and GSSP004-0011.

\bibliography{ms}
\clearpage \clearpage \thispagestyle{empty}

\begin{deluxetable}{llc} \tabletypesize{\scriptsize}  \tablewidth{9in}\rotate
\tablecaption{Sources of phase error in the VLBI experiment.\label{error}}
\tablehead{
     \colhead{Name}      &
     \colhead{Equation\tablenotemark{(a)}}   &
     \colhead{$\Delta \phi$\tablenotemark{(b)} } \\
     \colhead{}             &
     \colhead{} &
     \colhead{(degrees)}
}

\startdata

 Uncertainty in group delay estimate  &  $\displaystyle (2\pi
\Delta \nu)\left[\sqrt{\frac{3}{2\pi^2} }\left(\frac{T_S}{T_A}\right)
\frac{1}{\sqrt{\Delta \nu^3 t_{\rm cal}}} \right]$ & $0.5-1$ \\[15pt]

Residual delay error due to calibrator position error &  $\displaystyle (2\pi
\Delta \nu)\left[0.4\times 10^{-9}\,\frac{B}{5000\,\mbox{km}} \,\frac{\Delta
\theta_c}{5\,\mbox{mas}} \right]$ &
$0.07-0.2$ \\[15pt]

Residuals in atmospheric delay &  $(2\pi \Delta \nu)\Delta \sigma_{\rm atm}$ &
$<0.6$ \\[15pt]

Uncertainty in fringe rate estimate & $\displaystyle (2\pi \Delta
\nu)\,\frac{\Delta\omega}{\omega_o}\,t$ & $<0.5$ \\

Residual fringe rate from imperfect astrometry & $\displaystyle (2\pi \Delta
\nu)\,\left[0.13\times 10^{-3}\,\frac{B}{5000\,\mbox{km}} \,\frac{\Delta
\theta_c}{1\,\mbox{mas}} \right]\,\frac{t}{\omega_o}$
 & $0.3-1$ \\[15pt]

Errors in baseline length & $\displaystyle (2\pi \Delta \nu) \frac{\Delta B}{c}$
& $0.2$ \\[15pt]

Errors due to imperfect calibrator astrometry & $\displaystyle (2\pi \Delta \nu)
\frac{B \Delta \theta_p}{c}$ & $0.2$ \\[15pt]

Phase calibrator-target angular separation & $\displaystyle \frac{K}{\lambda_{\rm
mm}}
\,b_{km}^{\alpha}\,\mbox{deg}; \hspace{25pt}\displaystyle b_{km}=\left[\frac{h}{\sin e}\,\Delta \theta +\frac{v_at_{\rm cyc}}{2}\right]$ &  $<40$ \\
& where $\alpha=\frac{5}{6}$ for $b_{\rm km}<1.2$   \\[15pt]

\enddata

\tablenotetext{(a)}{Adopted from \cite{HerrnsteinPhDT} except for the last entry
which is based on \cite{Carilli1999}. $\Delta \nu=8$\,MHz is the video bandwidth
(i.e., the bandwidth across which various calibrations are applied and over which
the resulting errors are propagated), $T_{S}\sim1000$\,Jy is a representative
system equivalent flux density of VLBA antennas, $T_A=3-7$\,Jy is the flux
density of delay/rate calibrators (i.e., 0727-115, 0748+126, 4C\,39.25), $t_{\rm
cal}\sim3.5$\,min is the delay/rate calibrator scan duration, $\Delta
\theta_c=0.3-1$\,mas is the uncertainty in the delay/rate calibrator position \citep{Ma1998}, $B\sim5000$\,km is an approximate baseline length, $\Delta \sigma_{\rm atm}<0.2$\,nsec is the residual multiband delay after correcting for the zenith path delay over each antenna, $\Delta\omega<1$\,mHz is the residual rate after correcting for time variation in delays and fringe rates due to the troposphere and clock uncertainties, $t=1.5$\,hrs is an approximate time separation between the delay/rate calibrator scans, $\omega_o=22$\,GHz is the observing frequency, $\Delta B\sim2$\,cm is a representative uncertainty in baseline length, $c$ is the speed of light, $\Delta \theta_p=0.86$\,mas is the uncertainty in the phase calibrator (1055-248) position \citep{Beasley2002}, $b_{\rm km}$ is an effective baseline length in km, $K=200-600$ is a constant that depends on weather conditions (values assumed here are for the VLA site), $v_a<15$\,m\,s$^{-1}$ is the tropospheric pattern speed, $h\sim2$\,km is the water vapor scale height, $e\sim25\degr$ is a representative antenna elevation, $\Delta \theta=2.4\degr$ is the angular separation between the phase calibrator and the target source, $t_{\rm cyc}\sim100$\,s is the time between successive phase calibrator observations, and $\lambda_{\rm mm}=13$ is the observing wavelength in mm.}

\tablenotetext{(b)}{Phase error. On a 5000 km baseline, a phase error of 1 rad corresponds to a position error of $\sim 0.08$\,mas.}

\end{deluxetable}

\begin{deluxetable}{lccc}
\tablecaption{Velocities, positions, and integrated fluxes for mapped maser
emission.\label{maser_spots}} \tablehead{
     \colhead{Velocity\tablenotemark{(a)}}      &
     \colhead{RA\tablenotemark{(b)}}   &
     \colhead{DEC\tablenotemark{(b)}}  &
     \colhead{Flux\tablenotemark{(c)}}  \\
     \colhead{(km s$^{-1}$)} &
     \colhead{(mas)} &
     \colhead{(mas)} &
     \colhead{(Jy)}
}

\startdata

$3154.3$---$3157.7$ & $-0.3 \pm 0.2$ & $-0.9 \pm 0.4$ & $0.04 \pm 0.01$ \\
$3157.7$---$3161.1$ & $-0.5 \pm 0.3$ & $-0.4 \pm 0.7$ & $0.04 \pm 0.02$ \\
$3161.1$---$3164.6$ & $-0.3 \pm 0.2$ & $-0.4 \pm 0.4$ & $0.04 \pm 0.01$ \\
$3199.0$---$3202.5$ & $-0.3 \pm 0.2$ & $-0.5 \pm 0.5$ & $0.04 \pm 0.01$ \\
$3202.5$---$3205.9$ & $-0.1 \pm 0.2$ & $-1.7 \pm 0.4$ & $0.014 \pm 0.008$ \\
$3205.9$---$3209.4$ & $0.0 \pm 0.2$ & $-2 \pm 1$ & $0.03 \pm 0.01$ \\
$3223.2$---$3226.6$ & $0.1 \pm 0.1$ & $-1.5 \pm 0.3$ & $0.015 \pm 0.006$ \\
$3727.1$---$3730.5$ & $-0.8 \pm 0.2$ & $1.0 \pm 0.6$ & $0.012 \pm 0.009$ \\
$3730.5$---$3734.0$ & $-1.1 \pm 0.3$ & $0.3 \pm 0.6$ & $0.02 \pm 0.01$ \\
$3771.6$---$3775.0$ & $-1.3 \pm 0.2$ & $-0.2 \pm 0.3$ & $0.02 \pm 0.02$ \\
$4255.1$---$4258.6$ & $-2.5 \pm 0.2$ & $2.3 \pm 0.7$ & $0.04 \pm 0.01$ \\
$4258.6$---$4262.0$ & $-2.1 \pm 0.2$ & $1.3 \pm 0.5$ & $0.03 \pm 0.01$ \\
$4267.3$---$4270.7$ & $-2.5 \pm 0.1$ & $1.9 \pm 0.6$ & $0.02 \pm 0.01$ \\
$4270.8$---$4274.2$ & $-2.5 \pm 0.2$ & $1.5 \pm 0.3$ & $0.02 \pm 0.01$ \\
$4305.5$---$4308.9$ & $-2.3 \pm 0.2$ & $1.2 \pm 0.5$ & $0.02 \pm 0.01$ \\
$4308.9$---$4312.4$ & $-2.0 \pm 0.2$ & $1.0 \pm 0.6$ & $0.03 \pm 0.02$ \\
$4312.4$---$4315.9$ & $-2.0 \pm 0.1$ & $1.2 \pm 0.4$ & $0.06 \pm 0.02$ \\
$4315.9$---$4319.3$ & $-2.13 \pm 0.09$ & $1.6 \pm 0.2$ & $0.05 \pm 0.01$ \\
$4319.4$---$4322.8$ & $-2.20 \pm 0.07$ & $1.6 \pm 0.2$ & $0.06 \pm 0.01$ \\
$4322.8$---$4326.3$ & $-1.9 \pm 0.1$ & $1.8 \pm 0.4$ & $0.03 \pm 0.01$ \\

\enddata

\tablenotetext{(a)}{Range of mapped optical heliocentric velocities.}

\tablenotetext{(b)}{Right ascension and declination relative to
$\alpha_{2000}=10^h48^m23\fs4660$ and $\delta_{2000}=-25\arcdeg09\arcmin43\farcs478$, which lies $\sim1$\,mas from the estimated dynamical center (Section 4). The {\it a priori} position for the maser was $\alpha_{2000}=10^h48^m23\fs45$ and $\delta_{2000}=-25\arcdeg09\arcmin43\farcs6$ with uncertainty of $\pm0\rlap{.}''3$ from \cite{Kondratko2006}.}

\tablenotetext{(c)}{Integrated flux from a fit of a two-dimensional elliptical
Gaussian model to the distribution of the maser emission on the sky in each
spectral channel.}

\end{deluxetable}

\begin{deluxetable}{llcc@{\hspace{35pt}}cl}
\tablecaption{Fitted Line Profiles for Spectral Features Near 
$3880$\,km\,s$^{-1}$ and $4051$\,km\,s$^{-1}$.\label{gauss_fit_table}} \tablehead{
     \colhead{$\chi^2_R$\tablenotemark{(a)}}      &
     \colhead{Velocity\tablenotemark{(b)}}   &
     \colhead{Drift\tablenotemark{(c)}}  &
     \colhead{Amplitude\tablenotemark{(d)}}  &
     \colhead{FWHM\tablenotemark{(e)}}  &
     \colhead{Date}  \\
     \colhead{}             &
     \colhead{(km s$^{-1}$)} &
     \colhead{(km s$^{-1}$ yr$^{-1}$)} &
     \colhead{(mJy)} &
     \colhead{(km s$^{-1}$)}
}

\startdata

1.2 & $3871.0\pm 0.9$ & $5.3\pm0.7$ & \dots  & \dots &2005 Jan 15 \\
    &            &             & $5\pm 2$  & $4\pm 1$ &2005 Oct 17   \\
    &            &             & $10\pm 2$ & $2.6\pm 0.6$ &2006 Jan 22\\
    &            &             & $9\pm 2$  & $4.1\pm 0.9$ &2006 Mar 23\\
    &            &             & $10\pm 1$ & $5.2\pm 0.8$ &2006 Apr 28\\
    &            &             & $9\pm 2$  & $3.7\pm 0.9$ &2006 May 23\\
    & $3874.9\pm 0.2$ & $5.3\pm0.2$ & $ 6\pm 2$ & $3.6\pm 0.7$ &2005 Jan 15 \\
    &            &             & $41\pm 2$ & $3.4\pm 0.3$   &2005 Oct 17   \\
    &            &             & $43\pm 2$ & $3.3\pm 0.2$&2006 Jan 22\\
    &            &             & $32\pm 2$ & $3.2\pm 0.3$&2006 Mar 23\\
    &            &             & $29\pm 2$ & $2.4\pm 0.2$&2006 Apr 28\\
    &            &             & $15\pm 2$ & $4.2\pm 0.6$ &2006 May 23\\
    & $3878.9\pm 0.2$ & $4.4\pm0.2$ & $14.5\pm 0.9$ & $5.2\pm 0.5$   &2005 Jan 15 \\
    &            &             & $25\pm 5$ & $1.1\pm 0.3$    &2005 Oct 17   \\
    &            &             & $14\pm 2$ & $1.6\pm 0.3$&2006 Jan 22\\
    &            &             & $ 4\pm 2$ & $2.5\pm 0.9$&2006 Mar 23\\
    &            &             & $ 3\pm 1$ & $3\pm 1$&2006 Apr 28\\
    &            &             & \dots & \dots  &2006 May 23\\
    \tableline \\
2.0 & $4050.91\pm 0.01$ & $0.003\pm 0.009$ & \dots  & \dots    &2005 Oct 17   \\
    &            &              & \dots  &  \dots &2006 Jan 22\\
    &            &              & $119\pm 5$ & $0.99\pm 0.04$&2006 Mar 23\\
    &            &              & $300\pm 5$ & $0.80\pm 0.01$&2006 Apr 28\\
    &            &              & $254\pm 8$ & $0.80\pm 0.02$ &2006 May 23\\
    & $4051.9\pm 0.1$ & $-0.05\pm0.09$   & \dots    & \dots    &2005 Oct 17   \\
    &            &              & $13\pm 3$ & $1.4\pm 0.4$&2006 Jan 22\\
    &            &              & $11\pm 4$ & $1.4\pm 0.5$&2006 Mar 23\\
    &            &              & $33\pm 6$ & $0.7\pm 0.2$&2006 Apr 28\\
    &            &              & $23\pm 4$ & $1.2\pm 0.4$ &2006 May 23\\
    & $4053.06\pm 0.07$ & $-0.7\pm 0.6$ & $32\pm 5$ & $0.9\pm 0.2$  &2005 Oct 17   \\
    &            &              & $ 7\pm 3$ & $1.4\pm 0.8$&2006 Jan 22\\
    &            &              & \dots & \dots&2006 Mar 23\\
    &            &              & $ 6\pm 2$ & $1.4\pm 0.7$&2006 Apr 28\\
    &            &              & \dots & \dots  &2006 May 23\\

\enddata

\tablenotetext{(a)}{Reduced chi-squared for a least squares solution that
determines velocity drifts, amplitudes, peak centroids, and widths of all
Gaussian components at all epochs simultaneously.}

\tablenotetext{(b)}{Velocity of a Gaussian component at the reference epoch (2005
January 15 and 2005 October 17 for $\sim3880$\,km\,s$^{-1}$ and
$\sim4051$\,km\,s$^{-1}$ features, respectively).}

\tablenotetext{(c)}{Velocity drift of a Gaussian component.}

\tablenotetext{(d)}{Amplitudes of a Gaussian component at the listed epochs.}

\tablenotetext{(e)}{Full widths at half maximum of Gaussian components at the
listed epochs.}

\end{deluxetable}

\clearpage

\end{document}